\documentclass[twoside]{ilcws07}
\usepackage[latin1]{inputenc}
\usepackage[dvips]{graphicx,epsfig,color}
\usepackage{wrapfig,rotating}
\usepackage{amssymb,amsmath,array}

\def\mchio{\mbox{$m_{\chio}$}}
\def\ifmath#1{\relax\ifmmode #1\else $#1$\fi}%
\def\chio{\ifmath{\mathchoice%
     {\displaystyle\raise.4ex\hbox{$\displaystyle\tilde\chi{^0_1}$}}%
        {\textstyle\raise.4ex\hbox{$\textstyle\tilde\chi{^0_1}$}}%
      {\scriptstyle\raise.3ex\hbox{$\scriptstyle\tilde\chi{^0_1}$}}%
{\scriptscriptstyle\raise.3ex\hbox{$\scriptscriptstyle\tilde\chi{^0_1}$}}}}

\def\mt{\mbox{$m_{\rm \tilde{\rm t}_1}$}}
\def\mchio{\mbox{$m_{\chio}$}}
\def\st{\mbox{$\rm \tilde{t}_1$}}
\def\fb{\mathrm{\,fb}}

\newcommand{\mst}{m_{\tilde{t}_1}}

\newcommand{\anc}{\rule{0mm}{0mm}}
\newcommand{\mneu}[1]{m_{\tilde{\chi}^0_{#1}}}
\newcommand{\neu}{\tilde{\chi}^0}
\newcommand{\est}{\epsilon_{\rm \tilde{t}_1}}

\newcommand{\pt}        {p_T}

\begin{document}
\begin{titlepage}

\thispagestyle{empty}
\def\thefootnote{\fnsymbol{footnote}}       

\begin{center}
\mbox{ }

\end{center}

\vskip 3cm
\hspace*{-2cm}
\begin{picture}(0.001,0.001)(0,0)
\put(,0){
\begin{minipage}{1.2\textwidth}
\begin{center}
\vskip 1.0cm

{\Huge\bf
Scalar Top Studies
}
\vspace{3mm}

{\Huge\bf
from Morioka'95 to DESY'07
}
\vskip 1.5cm
{\LARGE\bf 
A. Sopczak$^1$,
A.~Finch$^1$,
A.~Freitas$^2$,
\vspace{3mm}

C.~Milst\'ene$^3$,
H.~Nowak$^4$,
M.~Schmitt$^5$

\bigskip
\bigskip

\Large $^1$Lancaster University, UK; $^2$Zurich University, Switzerland; \\
       $^4$Fermilab, USA; $^3$DESY, Germany; $^5$Northwestern University, USA}

\vskip 2cm
{\Large \bf Abstract}

\end{center}
\end{minipage}
}
\end{picture}

\vskip 6.5cm
\hspace*{-1cm}
\begin{picture}(0.001,0.001)(0,0)
\put(,0){
\begin{minipage}{\textwidth}
\Large
\renewcommand{\baselinestretch} {1.2}
Scalar top studies at the ILC are reviewed from initial sensitivity studies to 
a new precision mass determination method.
\renewcommand{\baselinestretch} {1.}

\normalsize
\vspace{5.5cm}
\begin{center}
{\sl \large
Presented at LCWS'07,
Linear Collider Workshop 2007 and the International Linear Collider meeting 2007,
DESY, Hamburg, Germany, 2007, \\
to be published in the proceedings.
\vspace{-6cm}
}
\end{center}
\end{minipage}
}
\end{picture}
\vfill

\end{titlepage}

\clearpage
\thispagestyle{empty}
\mbox{ }
\newpage
\setcounter{page}{1}
\pagestyle{plain}

\title{Scalar Top Studies from Morioka'95 to DESY'07}

\author{A. Sopczak$^1$\thanks{Email: andre.sopczak@cern.ch}, 
A.~Finch$^1$,
A.~Freitas$^2$,
H.~Nowak$^3$,
C.~Milst\'ene$^4$,
M.~Schmitt$^5$
\vspace{.3cm}\\
1- Lancaster U., 2- Zurich U., 3- DESY, 4- Fermilab, 5- Northwestern U.  \\
}

\maketitle

\begin{abstract}
Scalar top studies at the ILC are reviewed from initial sensitivity studies to 
a new precision mass determination method.
\end{abstract}

\vspace*{-4mm}
\section{Introduction}
\vspace*{-2mm}

Scalar top quarks have been studied in the framework of the ILC for  more than a decade.
In the following the developments since the International Linear Collider (ILC) 
workshop in Morioka 1995, where detection sensitivity was demonstrated, to recent 
precision mass determinations are presented.
The interplay with accelerator and detector aspects is \mbox{addressed} through the
importance of beam polarization for the accuracy of scalar top mass and mixing angle determination, 
and c-quark tagging for the vertex detector development. 
Different methods of scalar top mass determinations are addressed.
Particular attention is given to the scenario of small stop-neutralino mass differences.
The importance of scalar top studies at the ILC for the determination of the Cold Dark Matter (CDM) 
rate is emphasized. A new precision mass determination method, using two center-of-mass energies,
one near the production threshold, improves significantly the scalar top mass, 
as well as the CDM prediction.
The signal signature is two charm jets and missing energy from the process
$\rm e^+e^- \to \tilde{t}_1 \bar{\tilde{t}}_1 \to c \chio \bar c \chio .$

\vspace*{-2mm}
\section{Early Studies}
\vspace*{-2mm}
A detection sensitivity with more than about $7\sigma$ ($\sigma=S/\sqrt{B}$),
where $S$ is the number of expected signal and $B$ background events 
was demonstrated at the Linear Collider workshop in Morioka 1995~\cite{morioka},
as illustrated in Fig.~\ref{fig:morioka}.

\vspace*{-2mm}
\section{Developments from Morioka'95 to Sitges'99 to Jeju'02}
\begin{figure}[tp]
\begin{minipage}{0.75\textwidth}
\vspace*{-1cm}
\includegraphics[bbllx=316pt,bblly=293pt,bburx=524pt,bbury=504pt,clip=,angle=90,width=0.45\textwidth]{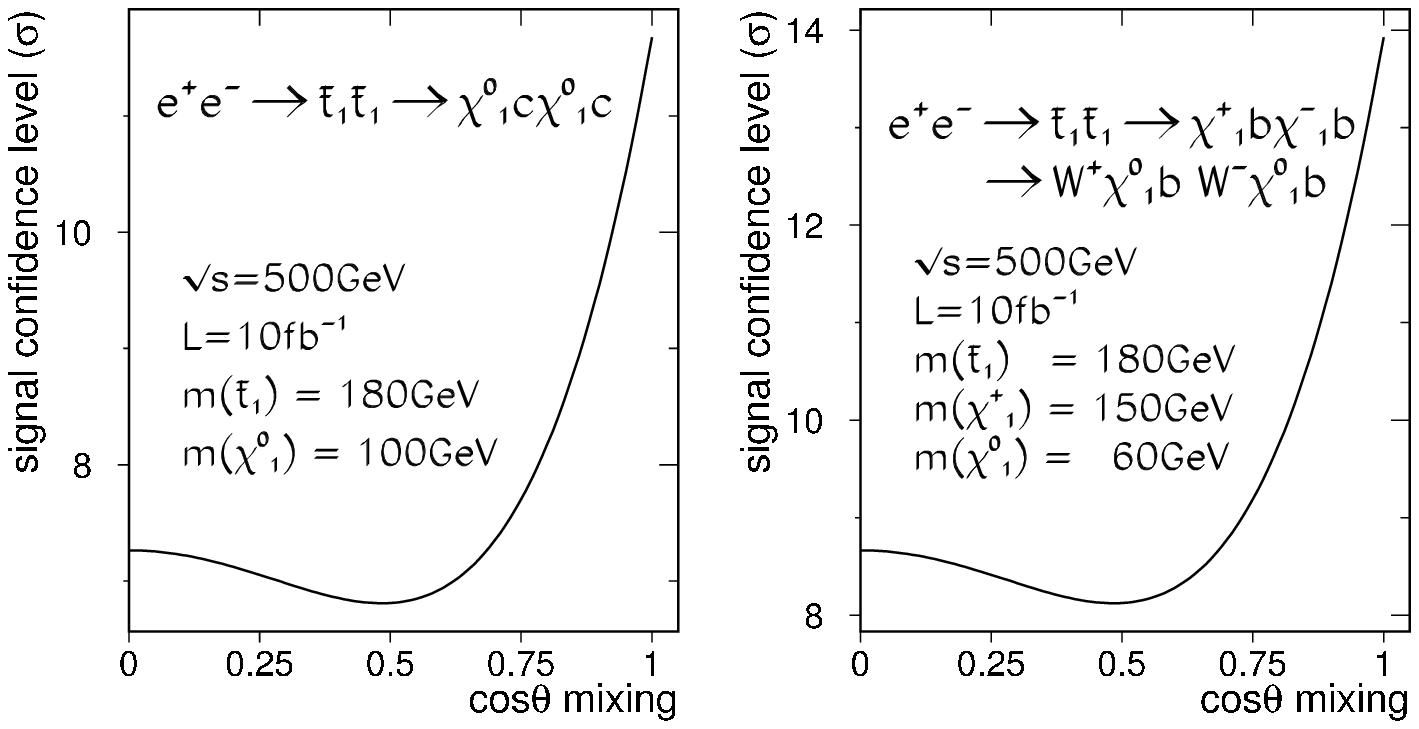}
\hfill
\includegraphics[width=0.55\textwidth]{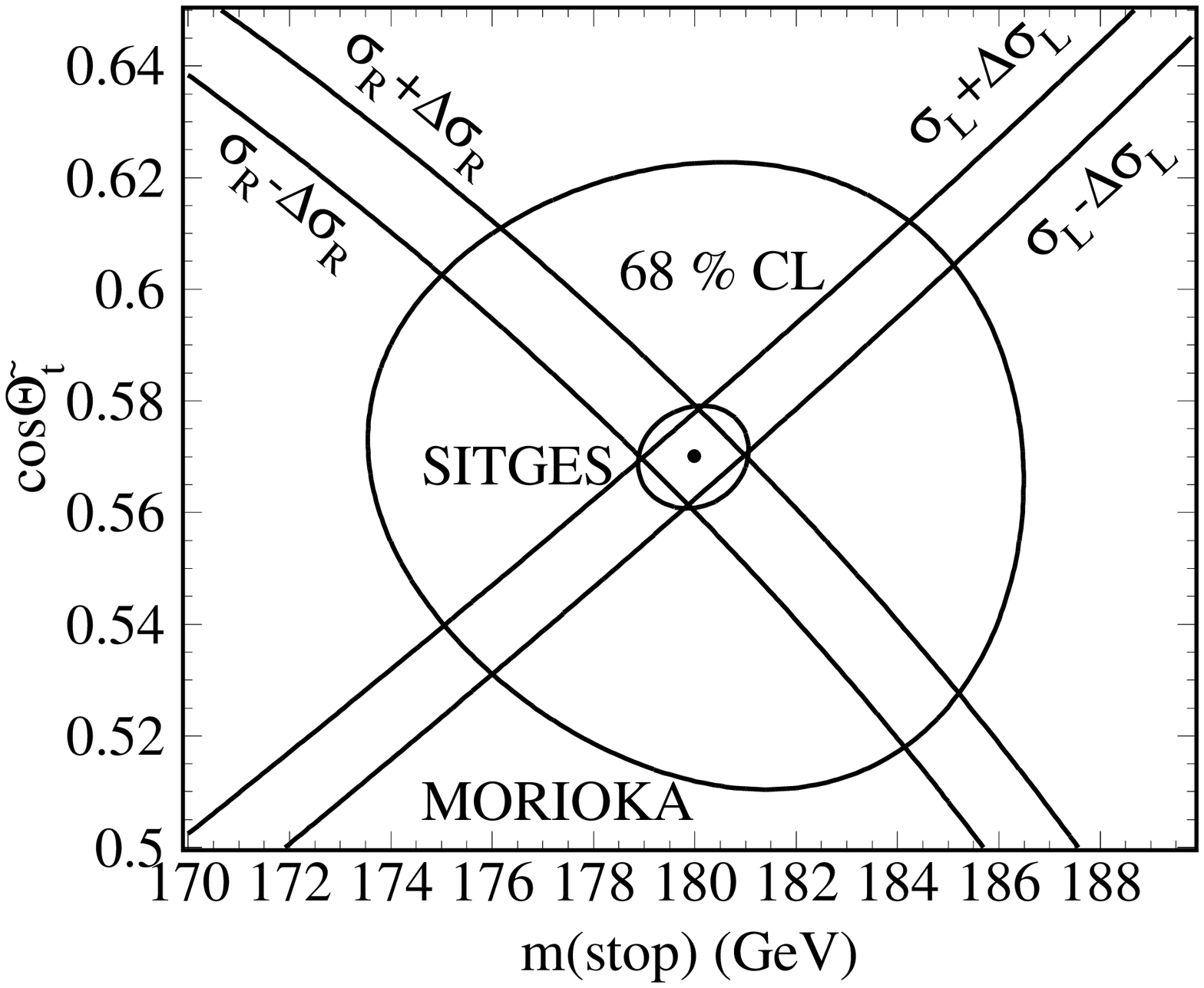}
\end{minipage}\hfill
\vspace*{-4mm}
\begin{minipage}{0.225\textwidth}
\caption{\small Left: initial sensitivity for scalar top quarks from the Morioka'95 workshop.
         Right: improvements of mass and mixing angle determination from Morioka'95 to Sitges'99.
\label{fig:morioka}
}
\end{minipage}
\vspace*{-2mm}
\end{figure}

\vspace*{-2mm}
At Morioka'95 the initial sensitivity was demonstrated for a luminosity of 10~fb$^{-1}$
and $\sqrt{s} = 500$~GeV using a LEP detector modeling.
Higher luminosities (500~fb$^{-1}$) have been assumed based on the accelerator developments
and presented at Sitges'99~\cite{sitges}.
In addition, an Iterative Discriminant (IDA) method was applied to separate expected signal and 
background events~\cite{sitges}.
Figure~\ref{fig:morioka} shows also the improvements in mass and mixing angle determination
($\mst=180.0\pm1.0$~GeV) including $\rm e^-$ beam polarization and the SGV detector modeling. 
Subsequently, the SIMDET detector description was used. Slightly higher precision was obtained 
in the neutralino channel including $\rm e^-$ and $\rm e^+$ beam polarization 
($\mst=180.0\pm0.8$~GeV), and 
the chargino decay mode was studied ($\mst=180.0\pm0.5$~GeV)~\cite{korea}.

\vspace*{-2mm}
\section{Major Challenge to Develop a Vertex Detector for the ILC}
\vspace*{-2mm}
During the LEP era (1989-2000 data-taking) the tagging of b-quarks with a vertex detector
was a major ingredient for many searches. The importance of c-quark tagging for scalar
top studies was realized. Key aspects are the distance between the interaction point and
the innermost layer of the vertex detector (radiation hardness, beam background) and the material absorption length 
(multiple scattering). A realistic vertex detector concept from the LCFI collaboration 
was implemented for c-quark tagging in the scalar top studies. 
Such a detector could consist of 5 CCD layers at 15, 26, 37, 48 and 60 mm, each layer 
with $<0.1\%$ absorption length.

\begin{figure}[htbp]
\vspace*{-0.3cm}
\begin{minipage}{0.49\textwidth}
\includegraphics[width=0.8\textwidth]{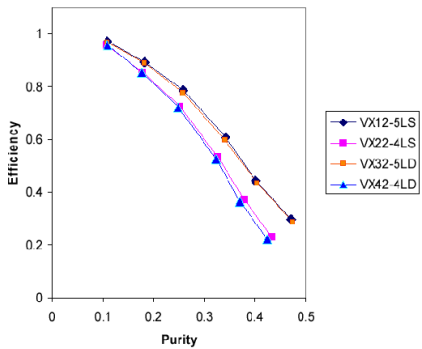}
\end{minipage} \hfill
\begin{minipage}{0.49\textwidth}
\caption{\small Scalar top c-tagging efficiency and purity with  $\rm W e\nu$ background 
for different detector designs. The VX$_{12}$ curve is for a detector design with 5 layers 
(innermost at 15~mm) and single density, curve VX$_{22}$ is for a detector design with 4
 layers (innermost at 26~mm). Curves VX$_{32}$ and VX$_{42}$ are for double density
(0.128\% absortion length per layer) 
with 4 and 5 layers, respectively.
} \label{fig:design}
\end{minipage}
\vspace*{-0.5cm}
\end{figure}

The importance of the vertex detector was studied with two different 
vertex detector configurations, one with 4 layers (removing the innermost layer), 
and the other one with 5 layers. The study was performed at $\sqrt{s}=500$~GeV
for a scenario with
large visible energy in the detector (Fig.~\ref{fig:design})~\cite{snowmass05} 
($\mt = 220.7$~GeV and $\mchio = 120$~GeV),
and one with small visible energy ($\mt = 122.5$~GeV and $\mchio = 107.2$~GeV) leading to
very similar results~\cite{bangalore06}. The innermost layer has a large effect on the 
c-tagging performance,
while doubling the detector thickness has a small effect.

\vspace*{-2mm}
\section{Signal Scenarios}
\vspace*{-2mm}
In order to investigate different detector scenarios and applying benchmark reactions 
for large and small visible energy three scalar top scenarios have been studied:
\begin{itemize}
\item For a comparison between different detector descriptions 
(SGV and SIMDET simulation packages) previous studies used $\mt = 180$\,GeV and $\mchio = 100$~GeV.
\item At the Snowmass'01 workshop the SPS-5 benchmark was established using MSSM parameters yielding 
      $\mt = 220.7$ GeV and $\mchio = 120$ GeV.

\item A cosmology motivated scenario has been studied in detail $\mt = 122.5$ GeV and $\mchio = 107.2$ GeV,
      including a sequential-cut-based analysis and using the IDA method.
\end{itemize}
In the first two scenarios the stop-neutralino mass difference is large and thus large visible 
energy is expected in the detector, while in the third scenario small visible energy is expected~\cite{carena}. 
The stop decay mode is always $\rm \st \rightarrow \tilde{\chi}^0_1 c$.

\vspace*{-2mm}
\section{Typical Analysis Strategy}
\vspace*{-2mm}
Since the study for the Jeju'02 workshop the basic analysis strategy remained unchanged
and signal and background processes have been generated for 500~fb$^{-1}$ and $\sqrt{s} = 500$~GeV.
A detector simulation (SIMDET) 
has been applied and a neural-network-based c-quark tagging algorithm 
has been used. The event selection has been performed with a sequential-cut-based analysis and an Iterative 
Discriminant Analysis (IDA).

\vspace*{-2mm}
\section{Four Different Methods of Mass Determination}
\begin{figure}[hp]
\begin{minipage}{0.58\textwidth}
\includegraphics[width=0.49\textwidth]{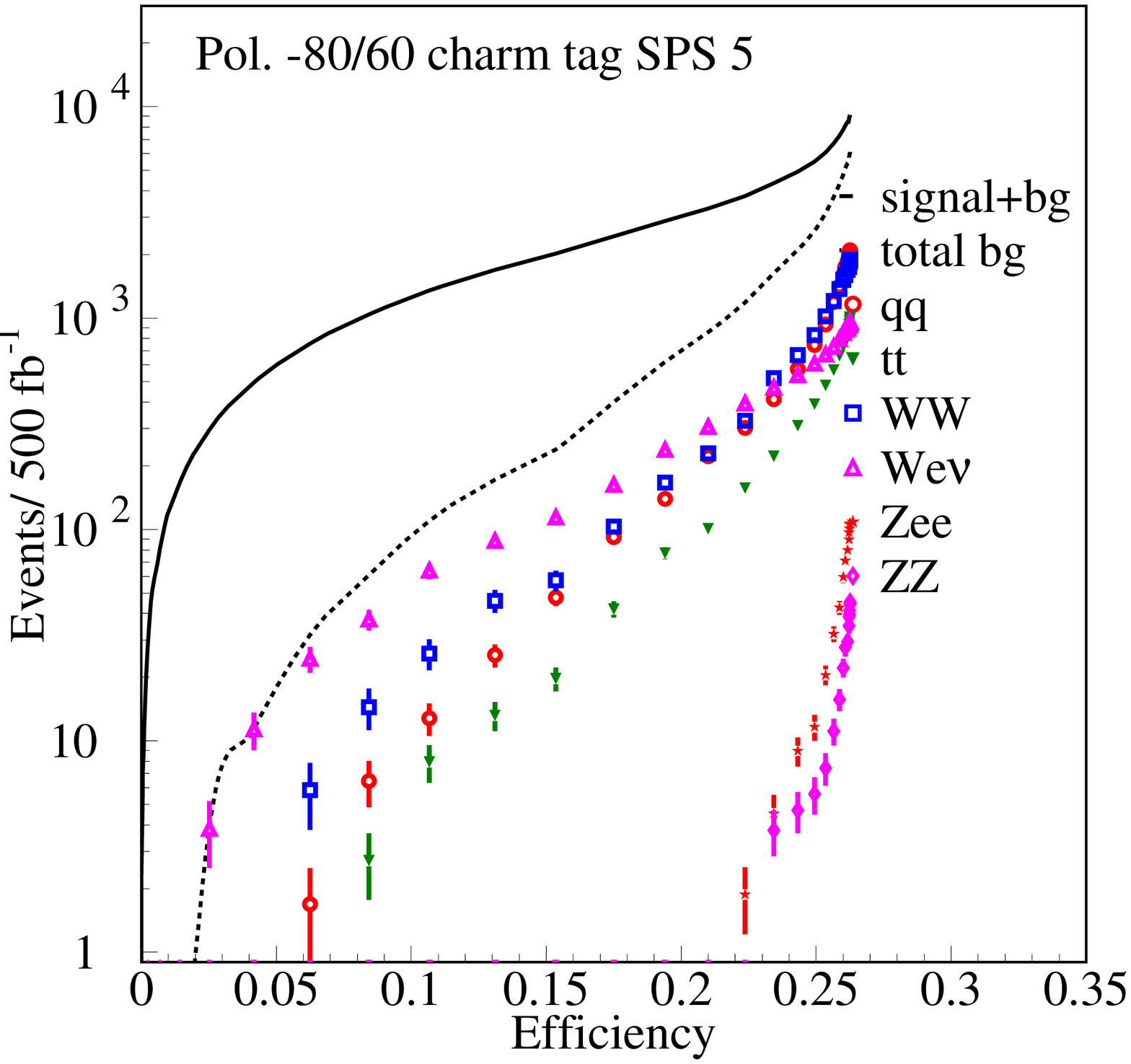}\hfill
\includegraphics[width=0.49\textwidth]{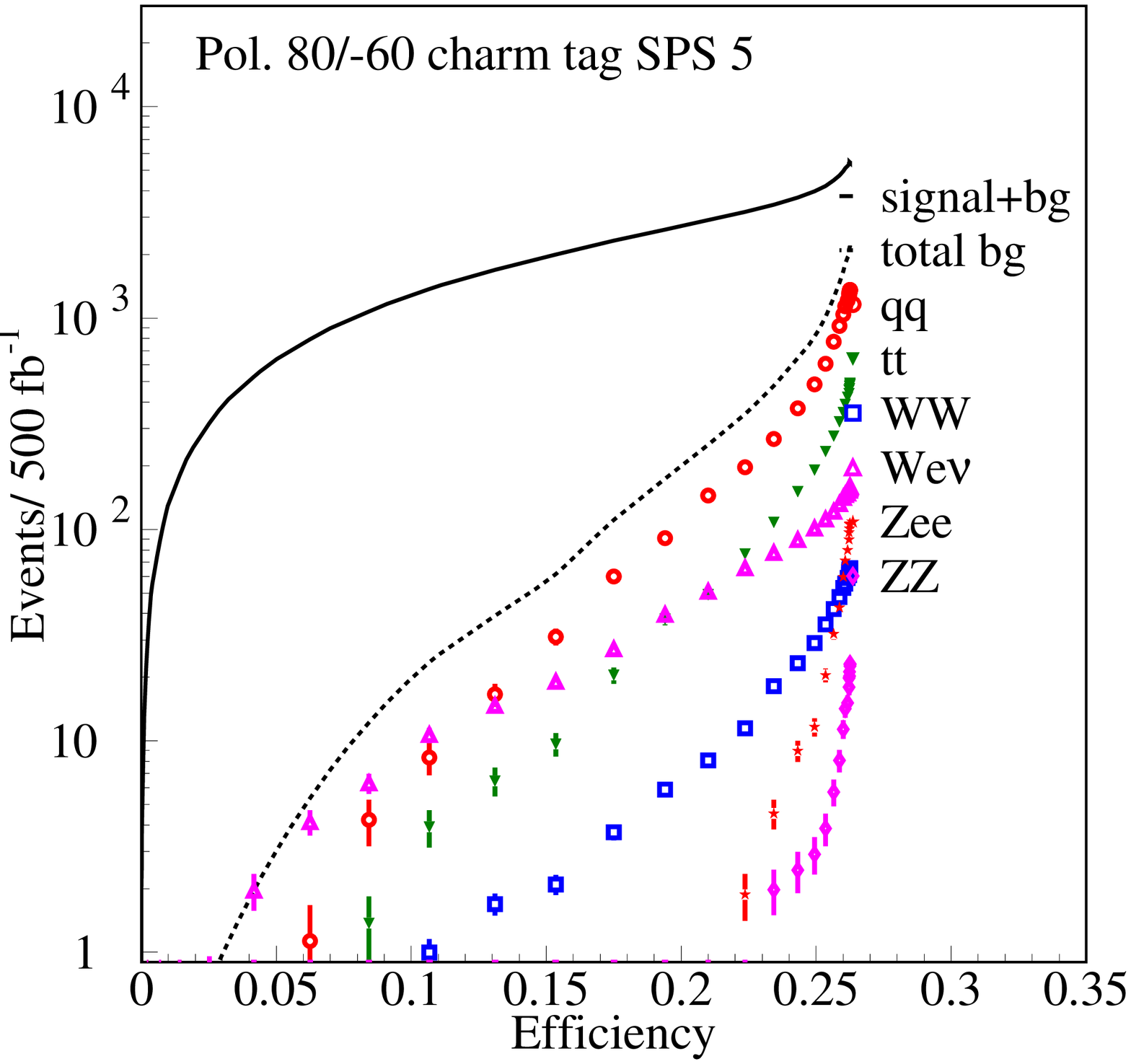}
\end{minipage}
\hfill
\begin{minipage}{0.39\textwidth}
\small
\begin{tabular}{lcc}\hline
Method & $\hspace*{-6mm} \Delta\mt$(GeV)\hspace*{-4mm}& luminosity \hspace*{-2mm}\\
\hline
Polarization & 0.57 & $2 \times 500 \mathrm{\fb}^{-1}$\hspace*{-2mm} \\
Threshold scan & 1.2 & $300 \fb^{-1}$ \\
End point & 1.7 & $500 \fb^{-1}$\\
Minimum mass & 1.5 & $500 \fb^{-1}$ \\ \hline
\end{tabular}
\vspace*{-4mm}
\caption{\small Left: expected number of background events as a function of the selection efficiency
         for two beam polarizations.
         Right: comparison of precision for scalar top mass determination
         for the SPS-5 benchmark.
\label{fig:sps5}
}
\end{minipage}
\vspace*{-4mm}
\end{figure}

\vspace*{-2mm}
Four different methods of mass determination were studied.
Two methods, which use the IDA for optimization of the signal to background ratio, are:
a) stop cross-section determination with different beam polarizations
(Fig.~\ref{fig:sps5}), and
b) threshold dependence of production cross-section.
Two cut-based selections were used in order to minimize the distortion of final state observables:
c) endpoint of jet energy spectrum, and 
d) minimum mass of jets.
These methods were discussed for the SPS-5 benchmark ($\mst=220.7$~GeV)~\cite{susy05}
and results are summarized also in Fig.~\ref{fig:sps5}. 

\vspace*{-4mm}
\section{Small Stop-Neutralino Mass Difference}
\begin{figure}[th]
\vspace*{-5mm}
\includegraphics[width=0.36\textwidth]{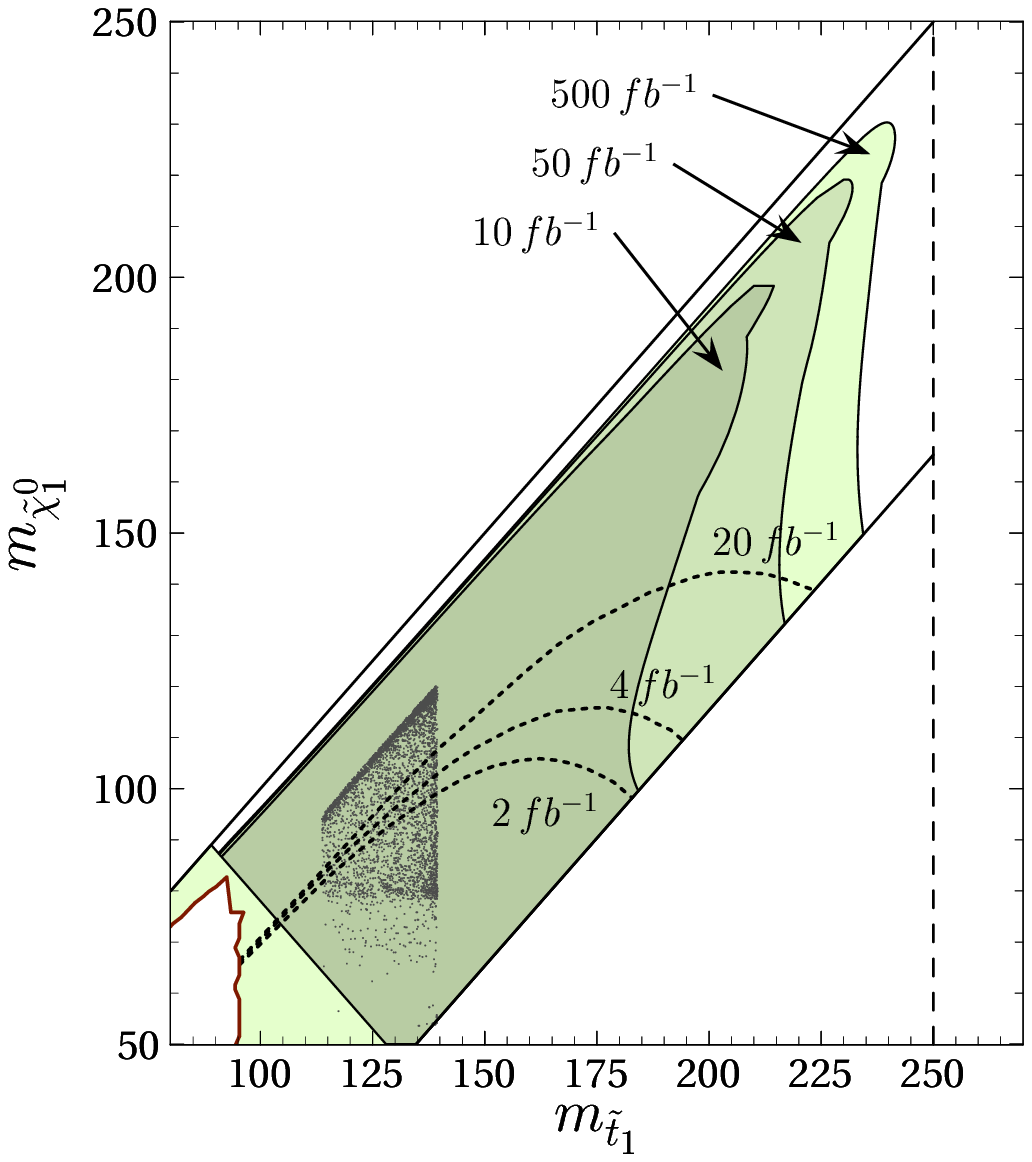} \hfill
\includegraphics[width=0.4\textwidth]{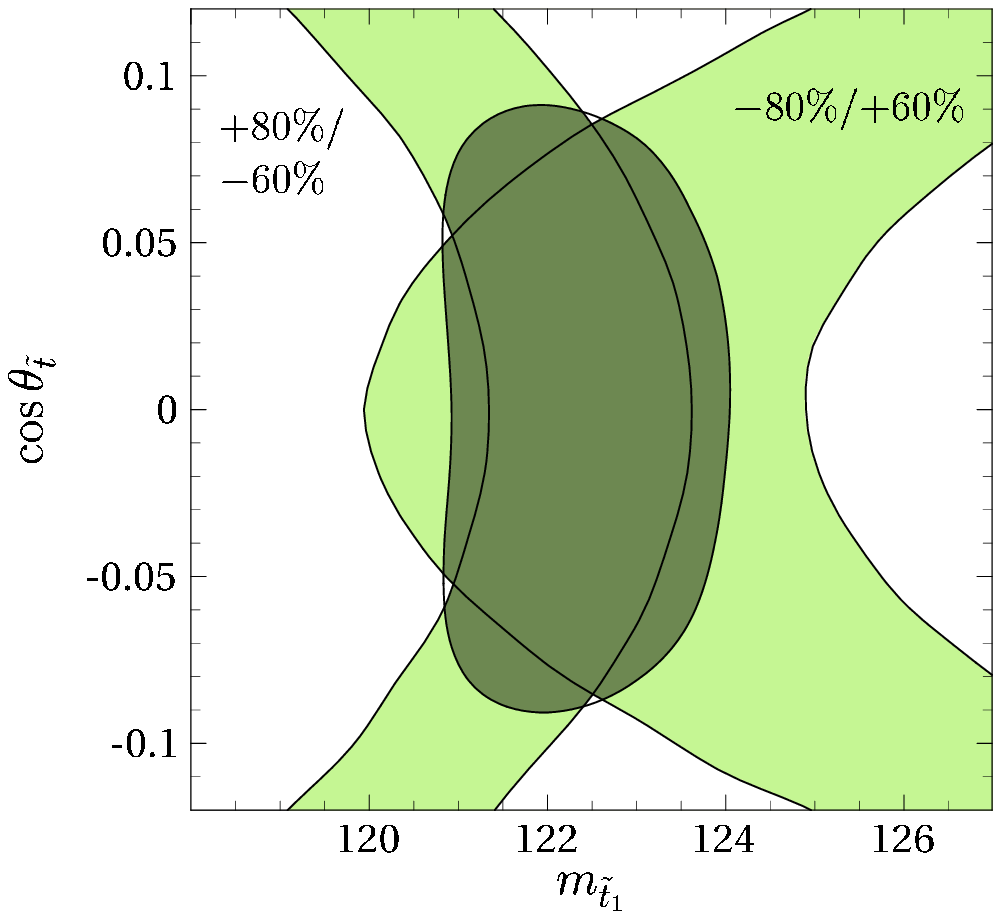}
\vspace*{-0.5cm}
\caption{
\small
Left: discovery reach of of the ILC with 10, 50, 500 fb$^{-1}$ luminosities at
$\sqrt{s} = 500$ GeV for the reaction 
$\rm e^+e^- \to \tilde{t}_1 \, \bar{\tilde{t}}_1 \to c \neu_1 \, \bar{c} \,\neu_1$. 
The results
are given in the stop vs. neutralino mass plane. In the gray shaded region, a
5$\sigma$ discovery is possible. The region $\mneu{1} > \mst$ is
inconsistent with a neutralino as Lightest Supersymmetric Particle (LSP), while for 
$\mst > m_{\rm W} + m_{\rm b} + m_{\neu_1}$ 
the three-body decay 
$\rm \tilde{t}_1 \to W^+ \bar{b} \neu_1$ becomes accessible and
dominant.  In the light shaded corner to the lower left, the decay of the top
quark into a light stop and neutralino is open.
The dark gray dots indicate the region consistent with baryogenesis and dark matter.
Also shown are the parameter
region excluded by LEP searches 
(white area in the lower left corner) and
the Tevatron light stop reach 
(dotted lines) for various integrated luminosities.
Right: Determination of light stop mass $\mst$ and stop mixing angle
$\theta_{\tilde{\rm t}}$ from measurements of the cross-section 
$\sigma(\rm e^+e^- \to \tilde{t}_1 \bar{\tilde{t}}_1)$ 
for beam polarizations $P({\rm e^-})/P({\rm e^+}) = -80\%/+60\%$ and
$+80\%/-60\%$. Statistical and systematic errors are included.
}
\label{fig:reach}
\vspace*{-2mm}
\end{figure}

\vspace*{-2mm}
A small stop-neutralino mass difference is motivated by cosmological aspects, 
baryogenesis 
$m_{\rm \tilde t_1} < m_{\rm t}$ and 
Dark Matter 
where $\tilde \chi^0_1$ is the\,Cold\,Dark\,Matter (CDM) candidate.\,A\,CDM  
rate consistent with observations is expected for a small 
$\tilde t_1 - \tilde\chi^0_1$ mass difference (co-annihilation).
The discovery reach is shown for $\mt = 122.5$ GeV and 
$\mchio = 107.2$ GeV~\cite{carena} (Fig.~\ref{fig:reach}).

For this small-mass-difference benchmark the determination of stop mass and mixing angle
were performed as for the previously described large mass difference scenario.
In the case of $\rm e^-$ and $\rm e^+$ polarization 
$\Delta m_{\tilde{\rm t}}=1.0$ GeV and 
$|\cos\theta_{\tilde{\rm t}} |<0.074$ was obtained, while for
$\rm e^-$ polarization only $\Delta m_{\tilde{\rm t}}=1.25$ GeV and 
$|\cos\theta_{\tilde{\rm t}} |<0.091$
was achieved~\cite{carena}.

For the CDM interpretation the following systematic uncertainties were taken into account:
$\Delta m_{\tilde\chi^0_1} =  0.3$~GeV,
polarization $\Delta P({\rm e^\pm}) / P({\rm e^\pm}) = 0.5\%$,
background rate $\Delta B / B = 0.3\%$,
scalar top hadronization and fragmentation ($<1\%$),
c-quark tagging ($<0.5\%$),
detector calibration ($<0.5\%$), and
beamstrahlung: ($<0.02\%$).
The sum of the systematic uncertainties is 1.3\% (left-handed beam polarization) and
1.2\% (right-handed beam polarization) without the theory error on the cross-section.
As the total systematic uncertainty is similar to the statistical uncertainty,
a reduction to $0.8\%$ was assumed being the same as the statistical uncertainty,
taking into account the LEP experience.
Including the expected theory uncertainty $\mt = 122.5 \pm 1.2 $ GeV was achieved.
The resulting CDM prediction included all parameters and their errors.
The stop mass uncertainty is dominant for the CDM co-annihilation precision.

\section{New Precision Mass Determination}
\vspace*{-2mm}
In order to improve the mass resolution, a new method has been proposed to measure 
the stop cross-section at two center-of-mass energies, one of them near the 
kinematic threshold where the cross-section is very sensitive to the stop mass,
and the other near the expected maximum production cross-section~\cite{ayres}.
The center-of-mass energies $\sqrt{s}= 260$~GeV and $\sqrt{s}= 500$~GeV are chosen.
This study also includes a more detailed description of the stop hadronization 
and fragmentation in the event simulation. Details are given in Ref~\cite{newmethod}.
For the event selection a sequential-cut-based analysis and the IDA method have been applied.

Both the sequential-cut-based analysis and the IDA method lead to small 
statistical uncertainties resulting in $\Delta \mst < 0.2$~GeV and thus
systematic uncertainties are particularly important to evaluate.
Four classes of systematic uncertainties are distinguished:
\begin{itemize}
\item instrumental uncertainties related to the detector and accelerator:
      detector calibration (energy scale),
      track reconstruction efficiency,
      charm-quark tagging efficiency, and
      integrated luminosity.
\item Monte Carlo modeling uncertainty of the signal: charm and stop fragmentation effects.
      The Peterson fragmentation function was used with
      $\epsilon_{\rm c}=-0.031\pm 0.011$ (OPAL). For 
      $\epsilon_{\rm b}=-0.0041\pm 0.0004$ (OPAL) and 
      $\epsilon_{\rm b}=-0.0031\pm 0.0006$ (ALEPH) an average uncertainty of 15\% was 
      taken, and a factor 2 improvement at the ILC has been assumed, 
      leading to $\Delta\est=0.6\times 10^{-6}$ where 
      $\est=\epsilon_b(m_{\rm b}/m_{\rm \mst})^2$.
      Fragmentation effects and gluon radiation increase the number of jets significantly and
      the importance of c-quark tagging is stressed in order to resolve the combinatorics.
\item neutralino mass $108.2\pm 0.3$~GeV.
\item theoretical uncertainties on the signal and background. Some improvement compared
      to the current loop calculation techniques is assumed, and an even larger reduction of this uncertainty
      is anticipated before the start of the ILC operation.
\end{itemize}

The systematic uncertainty
using the IDA method from detector calibration (energy scale) is large (Table~\ref{tab:sum}).
This is because the sequential-cut-based analysis pays particular attention to cancellation 
of this uncertainty between the two analyses at the different center-of-mass energies. 

The assessment of the achievable stop mass precision is based on the
statistical and systematic uncertainties on an observable $Y$ which is 
constructed from ratios of luminosities, selection efficiencies and 
theoretical production cross-sections at the two center-of-mass energies.
The IDA method has a smaller statistical uncertainty, and also a smaller background 
uncertainty due to a smaller number of expected background events.
The expected stop mass uncertainty is inferred from the uncertainty 
on $Y$ (Table~\ref{tab:sum}), as given in Table~\ref{tab:mstoperr}.

\begin{table}[hp]
\small
\begin{minipage}{0.74\textwidth}
\begin{tabular}{lcc}
\hline
\small
Error source for $Y$ & sequential cuts & IDA method \\
\hline
Statistical                           & 3.1\% &   2.7\% \\
Detector calibration                  & 0.9\% &   2.4\% \\
Charm  fragmentation                  & 0.6\% &   0.5\% \\
Stop fragmentation                    & 0.7\% &   0.7\% \\
Neutralino mass                       & 0.8\% &   2.2\% \\
Background estimate                   & 0.8\% &   0.1\% \\
\hline
Sum of experimental systematics       & 1.7\% &   3.4\% \\
Sum of experimental errors            & 3.5\% &   4.3\% \\
\hline
Theory for signal cross-section       & 5.5\% &   5.5\% \\
\hline
Total error $\Delta Y$                & 6.5\% &   7.0\% \\
\hline
\end{tabular}
\end{minipage}\hfill
\begin{minipage}{0.24\textwidth}
\caption{\small Summary of statistical and systematic uncertainties on
the observable~$Y$.
\label{tab:sum}
}
\end{minipage}
\end{table}

\begin{table}[hp]
\vspace*{-3mm}
\centering
\small
\begin{minipage}{0.74\textwidth}
\begin{tabular}{lcc}
\hline
 & \multicolumn{2}{c}{measurement error $\Delta\mst$ (GeV)} \\
Error category & sequential cuts & IDA method \\
\hline
Statistical                           & $0.19$   & $0.17$ \\
Sum of experimental systematics       & $0.10$   & $0.21$ \\
Beam spectrum and calibration         & $0.1\phantom{0}$    & $0.1\phantom{0}$  \\
Sum of experimental errors            & $0.24$   & $0.28$ \\
Sum of all exp. and th. errors
                                      & $0.42$   & $0.44$ \\
\hline
\end{tabular}
\end{minipage}\hfill
\begin{minipage}{0.24\textwidth}
\caption{\small Estimated measurement errors (in~GeV) on the stop quark mass.
\label{tab:mstoperr}}
\end{minipage}
\end{table}
\vspace*{-2mm}

\section{Cold Dark Matter (CDM) Interpretation}
\begin{figure}[tb]
\vspace*{-1cm}
\begin{center}
\begin{minipage}{0.45\textwidth}
\includegraphics[width=1\textwidth]{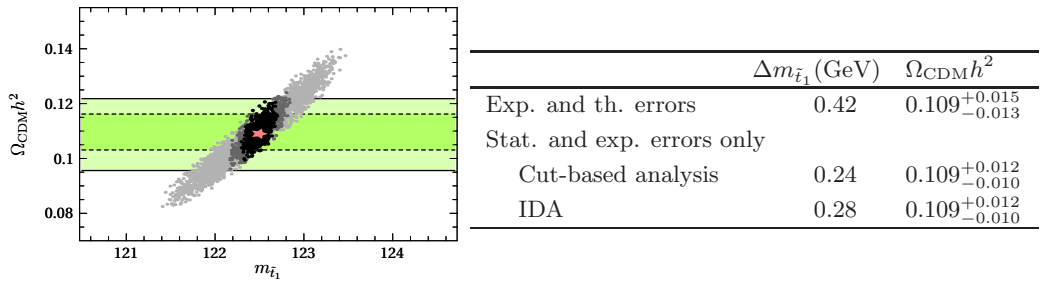}
\end{minipage} 
\hfill
\begin{minipage}{0.54\textwidth}
\renewcommand{\arraystretch}{1.2}
\small
\begin{tabular}{lcl}
\hline
 & \hspace*{-6mm}$\Delta \mst$(GeV)\hspace*{-1mm} & $\Omega_{\rm CDM} h^2$\hspace*{-3mm} \\
\hline
Exp. and th. errors   &
 0.42 & $0.109^{+0.015}_{-0.013}$ \\
Stat. and exp. errors only && \\ 
\anc\hspace{1em} Cut-based analysis &
 0.24 & $0.109^{+0.012}_{-0.010}$ \\
\anc\hspace{1em} IDA &
 0.28 & $0.109^{+0.012}_{-0.010}$ \\
\hline
\end{tabular}
\renewcommand{\arraystretch}{1}
\end{minipage}
\end{center}
\vspace*{-6mm}
\caption{\small
Left: expected dark matter relic abundance $\Omega_{\rm CDM} h^2$
taking into account detailed experimental errors for stop, chargino, neutralino
sector measurements at the future ILC. The black dots correspond to
a scan over the 1$\sigma$ ($\Delta \chi^2 \leq 1$) region including the 
total expected experimental uncertainties (detector and simulation),
the grey-dotted region includes also the theory uncertainty,
and the light grey-dotted area are the previous results~\cite{carena}.
The red star indicates the best-fit point. 
The horizontal shaded bands show the
1$\sigma$ and 2$\sigma$ constraints on the relic
density measured by WMAP.
Right:
estimated precision for the determination of stop mass and dark
matter relic density for different assumptions about the systematic errors.}
\label{fig:stoppar}
\vspace*{-0.2cm}
\end{figure}

\vspace*{-2mm}
The chosen benchmark parameters are compatible with the mechanism of electroweak
baryogenesis~\cite{carena}.
They correspond to a value for the dark matter relic abundance
within the WMAP bounds, $\Omega_{\rm CDM} h^2 = 0.109$.
The relic dark matter density has been computed as in Ref.~\cite{carena}\footnote{
The assumed benchmark parameters changed slighty (larger slepton masses assumed) and thus
$\Omega_{\rm CDM} h^2$ changed from 0.1122~\cite{carena} to 0.109.}.
In the investigated scenario, the stop and lightest neutralino masses are 
$m_{\rm \tilde{t}_1} = 122.5~$GeV and $\mneu{1} = 107.2$~GeV, and the stop mixing angle is 
almost completely right-chiral.
The improvement compared to Ref.~\cite{carena} regarding the CDM precision 
determination is shown in Fig.~\ref{fig:stoppar}~\cite{ayres}.

\section{\Large Conclusions }
\vspace*{-2mm}
Over the last decade the studies on scalar top quarks evolved from first
expected detection sensitivity (Morioka'95) to precision mass determination and
Cold Dark Matter predictions.
The $\rm e^-$ beam polarization is important for mass and mixing angle determination, and
the $\rm e^+$ polarization contributes in addition.
Detector simulations include c-quark tagging as a benchmark for vertex detector design studies.
Different detector descriptions (SIMDET and SGV) agree and 
dedicated simulations with SPS-5 parameters were performed.
Simulations for small stop-neutralino mass difference have been performed including hadronization
and fragmentation effects, leading to a larger number of jets.
An important aspect of this cosmology-motivated benchmark scenario is 
to resolve the jet-combinatorics by identifying the c-quark jets.
Precision mass determinations are possible with a method using two center-of-mass energies, 
e.g. $\sqrt{s}= 260$ and 500 GeV and
the expected ILC precision on $\Omega_{\rm CDM} h^2$ is comparable to WMAP measurements.
The ILC has a large potential to measure with precision scalar top quarks.
Scalar top quark studies have  addressed important questions related to 
accelerator and detector aspects. The proposed new method to measure the stop mass
with higher precision can also be applied to many other searches for new particles.

\end{document}